\documentclass[twoside]{article}

\usepackage{epsfig,fleqn,espcrc2}

\def\jou#1#2#3#4{{#1} {\bf #2} (#4) #3}

\def\NPA{{Nucl. Phys.} A}
\def\NPB{{Nucl. Phys.} B}
\def\PLB{{Phys. Lett.}  B}

\def\PRL{Phys. Rev. Lett.}

\def\PRD{{Phys. Rev.} D}

\def\EPJC{{Eur. Phys. J.} C}

\renewcommand\a{\alpha}
\renewcommand\c{\chi}
\newcommand\f{\phi}
\renewcommand\j{\psi}
\newcommand\p{\pi}

\newcommand\cm{{\cal M}}

\newcommand\dd{\mbox{d}}
\newcommand\K{{\mathbf k}}
\newcommand\intksno[1]{\frac{\dd^2 \K_{\perp\,#1}}{(2\pi)^2}}
\newcommand\Kpn[1]{\K_{\perp \,#1}}

\newcommand\lra{\longrightarrow}

\newcommand\be{\begin{equation}}
\newcommand\ee{\end{equation}}
\newcommand\bea{\begin{eqnarray}}
\newcommand\eea{\end{eqnarray}}

\newcommand\eref[1]{Eq.~(\ref{#1})}

\newcommand\fref[1]{Fig.~\ref{#1}}
\newcommand\tref[1]{Table~\ref{#1}}
\newcommand\bfi{\begin{figure}}
\newcommand\efi{\end{figure}}
\newcommand\bpi[1]{\begin{picture}#1}
\newcommand\epi{\end{picture}}

\begin{document}

\title{Color Octet Contribution in Exclusive P-Wave Charmonium Decay}

\author{S.M.H. Wong
\address{School of Physics and Astronomy, University of Minnesota, Minneapolis, 
Minnesota 55455, U.S.A.}
\thanks{The author thanks the organizing committee for all their works towards
this conference. This work was supported by the U.S. Department of Energy under 
grant no. DE-FG02-87ER40328.}
}

\begin{abstract}
Recent advances in our understanding of the higher-wave quarkonia have 
generated much interests in quarkonium physics. However most are devoted 
to inclusive decays and productions. Experimental data of several two-body 
exclusive decay channels of P-wave charmonia such as $\pi \pi$ and $p \bar p$ 
are available and some have recently been re-measured by the BES collaboration. 
It is not clear from the outset that color octet is needed for these exclusive 
channels. Indeed only color singlet has been used in the past and reasonable 
agreement with data was found. Contrary to these old results, we provide 
theoretical arguments for the inclusion of color octet and perform explicit 
calculations to back this up. %\hfill {\scriptsize NUC-MINN-00/13-T}
\end{abstract}

\maketitle

\section{Introduction}
\label{s:intro}

Quarkonia are special hadronic systems in that they have the 
mass as an intrinsic large scale. This permits factorization, 
the separation of the hard from the soft scale physics. Therefore 
once supplemented with non-perturbative matrix elements in 
inclusive processes or hadronic wavefunctions in exclusive 
processes, physical processes involving heavy quarkonia
can be calculated perturbatively. In this talk, we would like 
to re-examine some exclusive charmonium decay channels. There are 
several reasons for doing this. First being non-relativistic,
the heavy quark-antiquark system is simpler than most of the lighter 
hadrons and therefore is good for the study of color confinement in
this special case. If sufficiently understood, it provides a testing 
ground for our understanding of wavefunctions of selected lighter 
hadrons. Then recently several decay channels of P-wave charmonia 
have been measured or re-measured at the Beijing Spectrometer by
the BES collaboration so there are new data available. Third 
if one looks up the existing calculations, it is immediately evident
that they are rather out-of-date. Not only the QCD parameters
have changed, the hadronic wavefunctions used have also been
updated and better understood. Last but not least there is also the 
emergence of the color octet higher states in quarkonium, their
significance was revealed in \cite{bbl1,bbl2}. The simplest
decays are the two-body channels and so we will look at these
of the P-wave $\c_J$ system.

\section{Two-body P-wave charmonium decay}
\label{s:2bd}

A well-known scheme for calculating exclusive processes is the hard 
scattering approach (sHSA) of Brodsky and Lepage \cite{bl} which became a 
kind of standard so that is what the small ``s'' stands for in the acronym. 
This scheme relies on the presence of a large momentum transfer to validate
factorization so that the probability amplitude $\cm$ of the process in
question can be written as a convolution of the hadronic distribution
amplitudes $\f_h$ of the hadrons involved and a hard perturbative part
$T_H$ which serves as the conduit of the large momentum flow in the
Feynman diagram between the initial and final hadrons. For a two-body 
decay process resulting in hadron $h_1$ and $h_2$, this can be summarized 
in the following form 
\be   \cm \sim f_{\c_J} \f_{\c_J} \;\otimes\; T_H \;\otimes\; 
      f_{h_1} \f_{h_1} \;\otimes\; f_{h_2} \f_{h_2}  \; .
\ee  
The $f_h$'s are the decay constants of the respective hadrons. 
The convolution here is over the light-cone momentum fractions.
The partial decay width in terms of the above amplitude and the
heavy charmonium mass $M$ is essentially 
\be   \Gamma_{_{\c_J \rightarrow h_1 h_2}} \sim \frac{1}{M}\; |\cm|^2  \; .
\label{eq:wid}
\ee 
For definiteness we will examine the $\c_J$ decay into $\p \p$ and 
$p \bar p$. Both of these have been measured and data are available from
both Particle Data Group \cite{pdg} and BES \cite{bes}.  

The calculations of the decay into the above two channels are quite
reasonable with only two diagrams in the former and four in the latter
process and so are not too complicated \cite{bks,w1}. These are shown
in \fref{f:cs-pi} and \fref{f:cs-p}. 
\bfi
\centerline{\epsfig{figure=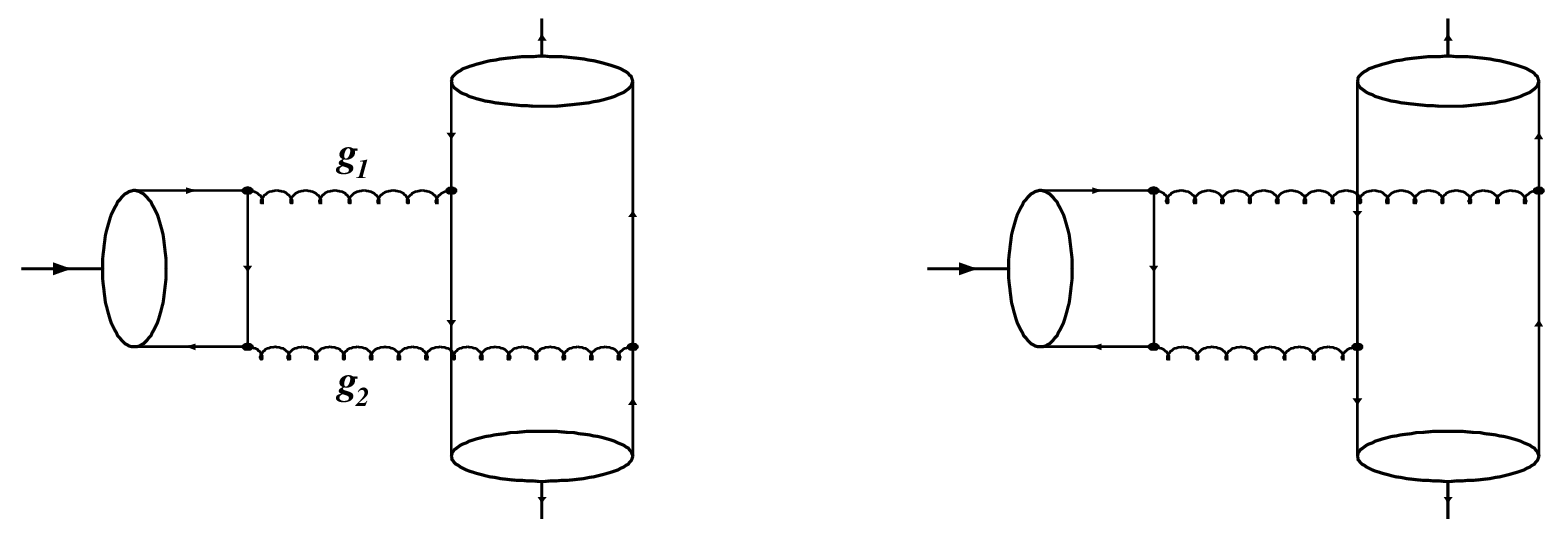,width=2.0in}}
\caption{The graphs for calculating the color singlet contribution for
the $\c_J$ decay into $\p\p$.}
\label{f:cs-pi}
\efi
\bfi
\centerline{\epsfig{figure=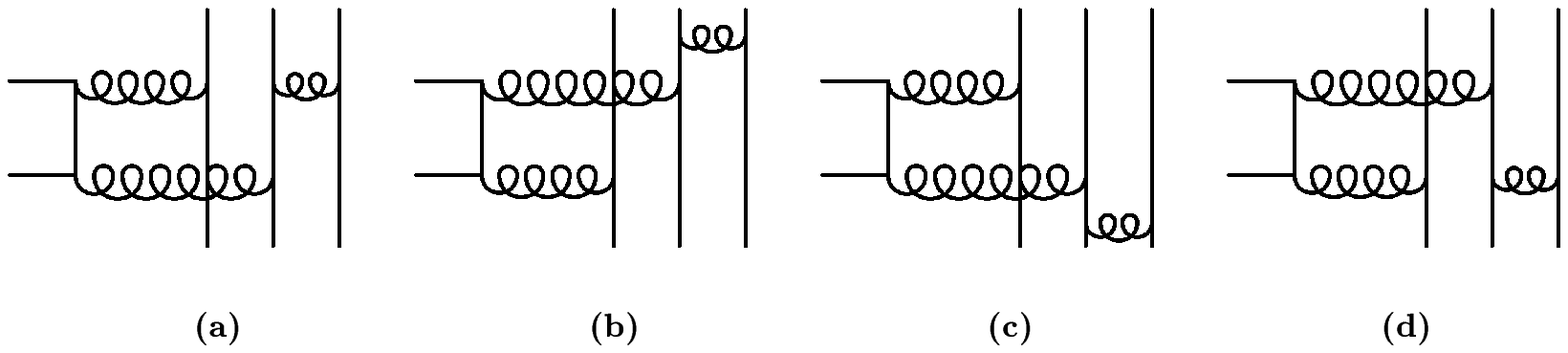,width=3.0in}}
\caption{The color singlet graphs for the decay into $p\bar p$.}
\label{f:cs-p}
\efi
Using the sHSA and some rather well-known quantities such as the $f_\p$, 
$\c_J$ radial wavefunction $R'_P(0)$, $m_c$ etc. the partial widths
can be obtained. One gets the results shown in \tref{t:r1}.
Comparing the partial widths $\Gamma^{(1)}$ coming from the $c\bar c$ 
state to those from experiments, one can see that the former are well 
below the data. 
\begin{flushleft}
\begin{table}
\caption{Partial decay widths obtained within the sHSA scheme.}
\begin{tabular}{cccc} \hline
 $J$  & \multicolumn{3}{c}{$\Gamma_{\c_J\rightarrow \pi\pi}$ [keV]} 
                                                                     \\ \cline{2-4}
      & \ $\Gamma^{(1)}$ \ \ & PDG \ \ & BES \ \                     \\ \hline
  0   &  15.30      & 105.0 $\pm$ 30   &  67.06 $\pm$ 17.3  \\
  2   &  0.841      &   3.8 $\pm$ 2.0  &   3.04 $\pm$ 0.73  \\ \hline 
\end{tabular}

\null 
\begin{tabular}{cccc} \hline
 $J$  & \multicolumn{3}{c}{$\Gamma_{\c_J\rightarrow p\bar p}$ [eV]} 
                                                                  \\ \cline{2-4}
      & \ $\Gamma^{(1)}$ \ \ & PDG \ \ & BES \ \                  \\ \hline
  1   &  3.15       &  75.68 $\pm$ 10.5  &   37.84 \ \ \ \    \\
  2   &  12.29      & 200.00 $\pm$ 20.0  &  118.00 \ \ \ \    \\ \hline 
\end{tabular} 
\label{t:r1}
\end{table} 
\end{flushleft}

\vspace{-0.9cm}
In order to make sure that these small values are not the result of 
the imperfection of the scheme used. They can be calculated again using 
an improved scheme (mHSA) of Sterman et al \cite{bs,ls}. This scheme 
comes from modifying the sHSA by introducing transverse momentum effects, 
radiative correction in the form of Sudakov factor $S$ and utilizing whole 
hadronic wavefunctions $\j_h$. The probability amplitude can now be 
summarized as 
\be \cm \sim \j_{\c_J} 
      \;\otimes\; T_H ({\a_s}) \;\otimes\; 
      \j_{h_1} \;\otimes\; \j_{h_2} \;\otimes\; \mbox{exp}\{-S\} \; .
\ee
The convolution is now over not just light-cone fractions but also
internal transverse momenta. Furthermore the coupling $\a_s$ is 
part of the convolution whose scale is determined dynamically by
the momentum flow, whereas in sHSA it is merely a constant. The
importance of this was discussed in \cite{w1,w2}. The results
shown in \tref{t:r2} agree essentially with those in \tref{t:r1} 
on the fact that the theoretical results are too small. Of course, 
one could exploit the uncertainties in the various parameters
to try to enlarge the calculated results but this proved to 
be fruitless. They remain below the data by factors of two or 
more \cite{bks,w1,bks2,w3}. 
\begin{flushleft}
\begin{table}
\caption{Partial decay widths obtained within the mHSA scheme.}
\begin{tabular}{cccc} \hline
 $J$  & \multicolumn{3}{c}{$\Gamma_{\c_J\rightarrow \pi\pi}$ [keV]} 
                                                                     \\ \cline{2-4}
      & \ $\Gamma^{(1)}$ \ \ & PDG \ \ & BES \ \                     \\ \hline
  0   &  8.22     & 105.0 $\pm$ 30   &  67.06 $\pm$ 17.3  \\
  2   &  0.41     &   3.8 $\pm$ 2.0  &   3.04 $\pm$ 0.73  \\ \hline 
\end{tabular} 

\null
\begin{tabular}{cccc} \hline
 $J$  & \multicolumn{3}{c}{$\Gamma_{\c_J\rightarrow p\bar p}$ [eV]} 
                                                                     \\ \cline{2-4}
      & \ $\Gamma^{(1)}$ \ \ & PDG \ \ & BES \ \                     \\ \hline
  1   &  2.53       &  75.68 $\pm$ 10.5  &   37.84 \ \ \ \ \         \\
  2   &  16.58      & 200.00 $\pm$ 20.0  &  118.00 \ \ \ \ \         \\ \hline 
\end{tabular}
\label{t:r2}
\end{table}
\end{flushleft}
\vspace{-0.730cm}

\section{What is the source of the problem?}
\label{s:sop}

The previous section showed that it was not possible to explain
the experimental data from the decay of the P-wave $c\bar c$ system.
In the search for the source of the problem, one cannot help but
wonder if color octet is the answer. After all color octet seems these 
days to be inextricably intertwined with quarkonium physics in inclusive
processes ever since it was first introduced in \cite{bbl1,bbl2} to cancel 
the infrared divergence found in the inclusive $\c_J$ decay \cite{bg}. 
Strangely in the world of exclusive quarkonium processes, color octet 
does not seem to exist at all. In fact it has been totally neglected. 
A simple argument easily reveals that color octet must also be important
in exclusive decays, albeit it is not sufficient to claim that it
is so in the two-body channels. The argument relies on the 
fact that the sum of the partial widths of all channels must be equal
to the total inclusive hadronic width which can be more precisely 
expressed in the form
\be  \sum_{i\in \mbox{\scriptsize channels}} \Gamma_i 
     = \Gamma_{\mbox{\scriptsize incl.}}  \; .
\ee
If none of the $\Gamma_i$ requires the color octet contribution, then
there would be a contradiction because it is known to be of importance and
needed on the right-hand-side (r.h.s). However the contradiction can be 
avoided provided color octet is needed in at least one channel, it need 
not be the two-body channels that we are interested in. So {\em a priori} 
for our problem at hand, it is not clear that color octet is the answer. 
Furthermore there is the usual folklore that says higher state 
contributions are suppressed in large momentum transfer processes
which seems to oppose the inclusion of the octet contribution.

\section{The large mass dependence of the decay amplitudes}
\label{s:lmd}

The pros and cons of color octet as the answer to our problem 
discussed in the last section makes the issue very unclear. 
To come to a resolution, it is necessary to look closely at the 
charmonium system. The hadronic decay is via annihilation into
gluons. The $c\bar c$ pair has to come close together for this
at a small distance $l \sim 1/M$ because of the large charmonium
mass $M$. For a P-wave charmonium, the $L=1$ orbital angular
momentum tends to force the pair apart therefore it is harder for 
the pair to annihilate. This is manifested in the vanishing
of the P-wave wavefunction at the origin $\psi_P(0)=0$. Comparing 
a S-wave to a P-wave system, assuming all else being equal, the 
only difference in magnitude of the decay amplitude lies entirely 
with the wavefunctions at small distance. Because the P-wave 
wavefunction vanishes there, its derivative enters the decay 
amplitude in its place. The following shows the relevant quantities 
that enter the amplitudes and the same after transforming into 
momentum space.  
\begin{eqnarray*}
   S:  & \j_S (l\sim 0)                          
       & \rightarrow  \tilde \j_S (k)              \\
   P:  & \j_P (l\sim 0) \simeq l \j'_P (l\sim 0) 
       & \rightarrow  \frac{k}{M} \tilde \j_P (k)  \\
\end{eqnarray*}
This shows the P-wave quantity carried with it a power of $1/M$
so a P-wave charmonium is suppressed at the valence level already
by $M$ in comparison to a S-wave. In view of this power suppression,
it is perhaps useful to examine and compare the $M$ dependence of the 
decay amplitudes. 

In the calculations of exclusive process, one has to deal all too
frequently with decay constants of the hadrons involved. So it 
should be more relevant to look at the decay amplitudes in terms
of $f_h$'s and $M$. The sHSA scheme is therefore more convenient for
this purpose. Since the strongest dependence on $M$ is powerlike,
it is useful to look at dimensions. From \eref{eq:wid} $\cm$
is of mass dimension one so we have the following equation of 
dimensions 
\be [\cm] = [{\rm mass}]^1 = [f_{\c_J}] [f_{h_1}] [f_{h_2}] [T_H] \; .
\ee
The r.h.s. shows all the quantities that carried a mass dimension. 
The hard part $T_H$ has some hidden power dependence of the form $M^{-p}$
for some number $p$. Together with the $f_h$'s, they make up for the
dimension of $\cm$. The $M$ dependence of $\cm$ will be known, once 
the mass dimensions of the $f_h$'s are determined. Using the 
relation of $f_h$ to the associated wavefunction $\j_h$ 
\bea & & f_h \f_h (x)                         \nonumber \\
     &=&\!\!\!\!\! \int^Q \prod^{N-1}_i \intksno{i} \; 
         \j_h (x;\Kpn{1},\dots,\Kpn{N-1})
\eea
and the normalization of $\j_h$
\bea & & \!\!\!\!\! \int \prod^{N-1}_i \dd x_i \intksno{i} \;
       \left | \j_h (x;\Kpn{1},\dots,\Kpn{N-1}) \right |^2 \nonumber \\
     & &\!\!\!\!\! = \mbox{prob.} \; ,
\eea
one can deduce the dimension of $f_h$. It can be summarized by  
\be [\j_h] = [{\rm mass}]^{1-N}  \;, \hspace{0.6cm} 
    [f_h]  = [{\rm mass}]^{N-1+L} 
\ee
where $N$ is the number of constituents in the hadron $h$ described 
by $\j_h$ and $L$ is the orbital angular momentum of $\j_h$. 
The decay constants for our hadrons are then
\bea [f_\p] = [{\rm mass}]^1, &&
     [f_p] = [f^{(1)}_{\c_J}] = [f^{(8)}_{\c_J}] =[{\rm mass}]^2 
     \; . \nonumber 
\eea 

For the $\c_J$ decays, the color singlet and octet amplitude 
therefore have the following $M$ dependence
\begin{eqnarray*}
     \cm^{(1)}_{\c_J \lra \pi \pi} & \sim & 
     M \frac{f^{(1)}_{\c_J}}{M^2} \Big ( \frac{f_\pi}{M} \Big )^2  
     \sim \frac{1}{M^{3}}   \\
     \cm^{(8)}_{\c_J \lra \pi \pi} & \sim & 
     M \frac{f^{(8)}_{\c_J}}{M^2} \Big ( \frac{f_\pi}{M} \Big )^2  
     \sim \frac{1}{M^{3}} 
\end{eqnarray*}
\begin{eqnarray*}
     \cm^{(1)}_{\c_J \lra p\bar p} & \sim & 
     M \frac{f^{(1)}_{\c_J}}{M^2} \Big ( \frac{f_p}{M^2} \Big )^2  
     \sim \frac{1}{M^{5}}   \\
     \cm^{(8)}_{\c_J \lra p\bar p} & \sim & 
     M \frac{f^{(8)}_{\c_J}}{M^2} \Big ( \frac{f_p}{M^2} \Big )^2  
     \sim \frac{1}{M^{5}}   \; .
\end{eqnarray*}
For both decay channels, singlet and octet contribution have the
same power dependence in $M$. There is no large $M$ suppression
of the octet in comparison to the singlet. The usual suppression of
the higher state is nullified by the suppression of the P-wave
valence state due to angular momentum. 

To elucidate this further, we can turn to the same decay channels of 
the S-wave $J/\j$. Now the $J/\j$ decay constants have the dimensions
\bea && [f^{(1)}_{J/\j}] = [{\rm mass}]^1, \hspace{0.6cm}
        [f^{(8)}_{J/\j}] = [{\rm mass}]^3  \; . \nonumber 
\eea 
The octet constant has a dimension of three instead of two is because 
the three-body wavefunction must have $L=1$ to make up the right
quantum numbers for $J/\j$ as explained in \cite{w1}. The $J/\j$
decay amplitudes have the dependence
\begin{eqnarray*}
     \cm^{(1)}_{J/\j \lra \pi \pi} & \sim & 
     M \frac{f^{(1)}_{J/\j}}{M} \Big ( \frac{f_\pi}{M} \Big )^2  
     \sim \frac{1}{M^{2}}        \\
     \cm^{(8)}_{J/\j \lra \pi \pi} & \sim & 
     M \frac{f^{(8)}_{J/\j}}{M^3} \Big ( \frac{f_\pi}{M} \Big )^2  
     \sim \frac{1}{M^{4}}   
\end{eqnarray*}
\begin{eqnarray*}
      \cm^{(1)}_{J/\j \lra p\bar p} & \sim & 
     M \frac{f^{(1)}_{J/\j}}{M} \Big ( \frac{f_p}{M^2} \Big )^2  
     \sim \frac{1}{M^{4}}       \\
     \cm^{(8)}_{J/\j \lra p\bar p} & \sim & 
     M \frac{f^{(8)}_{J/\j}}{M^3} \Big ( \frac{f_p}{M^2} \Big )^2  
     \sim \frac{1}{M^{6}}  \; .
\end{eqnarray*}
So the octet contribution is suppressed by $M^{-2}$ in both cases
in the amplitude. One can therefore legitimately neglect the color 
octet contributions in $J/\j$ but not in $\c_J$.

\section{Color octet contributions}
\label{s:co}

In this section, we briefly outline and discuss the calculation
of the octet contributions. The calculation can be done in either sHSA or 
mHSA scheme in principle, in practice it is easier to use sHSA than 
mHSA but the latter has important advantages discussed in \cite{w1,w2,w3,w4}. 
To some degree the choice of scheme depends also on the final hadrons. 
The simpler $\p \p$ is manageable in mHSA but not the $p\bar p$
channel which is too complicated. It is better to work in sHSA 
for calculating the latter. 

Since the color octet wavefunctions 
are not known, they have to be constructed. To do this the 
associated decay constants $f_{\c_J}^{(8)}$ must be determined by
fitting and the light-cone fractions' distribution fixed in a sensible
manner. Numerically the latter has to be checked to see if it is sensible. 
Then there is the presence of the constituent gluon from the octet
state of the charmonium which has to be considered. The usual 
valence color singlet $c\bar c$ state is straight forward 
to deal with since they cannot appear in the final state and 
easily eliminated via annihilation. The constituent gluon on the
other hand can be a part of the final hadron by becoming a constituent
of a higher state or it can end in the hard perturbative part
$T_H$. It was eventually decided that the dominant contribution 
would be for it to end on $T_H$. The contribution involving the
higher state of a final hadron would be suppressed both by large 
momentum flow within HSA and the presence of the additional
higher state. To complete the calculation of $T_H$, all possible
feynman graphs must be found. Unlike the color singlet contribution,
there are now a lot more diagrams. Because of C-parity, the color
singlet state must annihilate into two or more gluons. The $c\bar c$
in the color octet on the other hand is protected from the C-parity
constraint by the presence of the constituent gluon and can therefore
annihilate through one gluon. This leads to quite a number of diagrams.
Additionally by allowing the constituent gluon to end in the hard part, 
it must be attached to all possible allowed positions of $T_H$. 
These all contribute to the number of graphs. In the end, they can
be organized into groups and calculated by computer. Some examples
are shown in \fref{f:pi} and \fref{f:p}. The first figure here 
is from \cite{bks}. The constituent gluon is yet to be 
included in the latter. The graphs shown in this one 
form the basis from which the groups are to be generated. 
\bfi[t]
\centerline{\epsfig{file=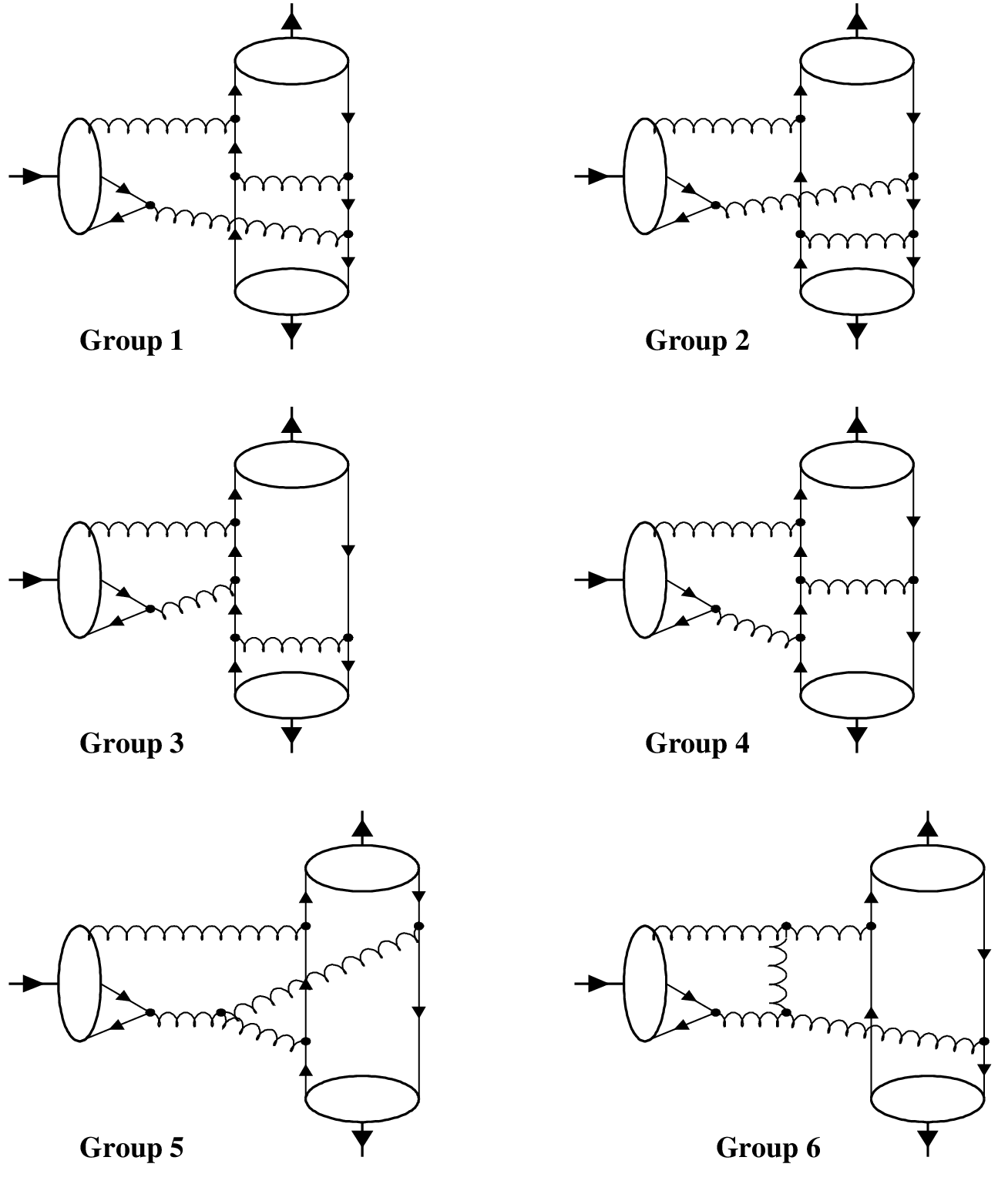,width=3.70cm} \hspace{0.3cm}
            \epsfig{file=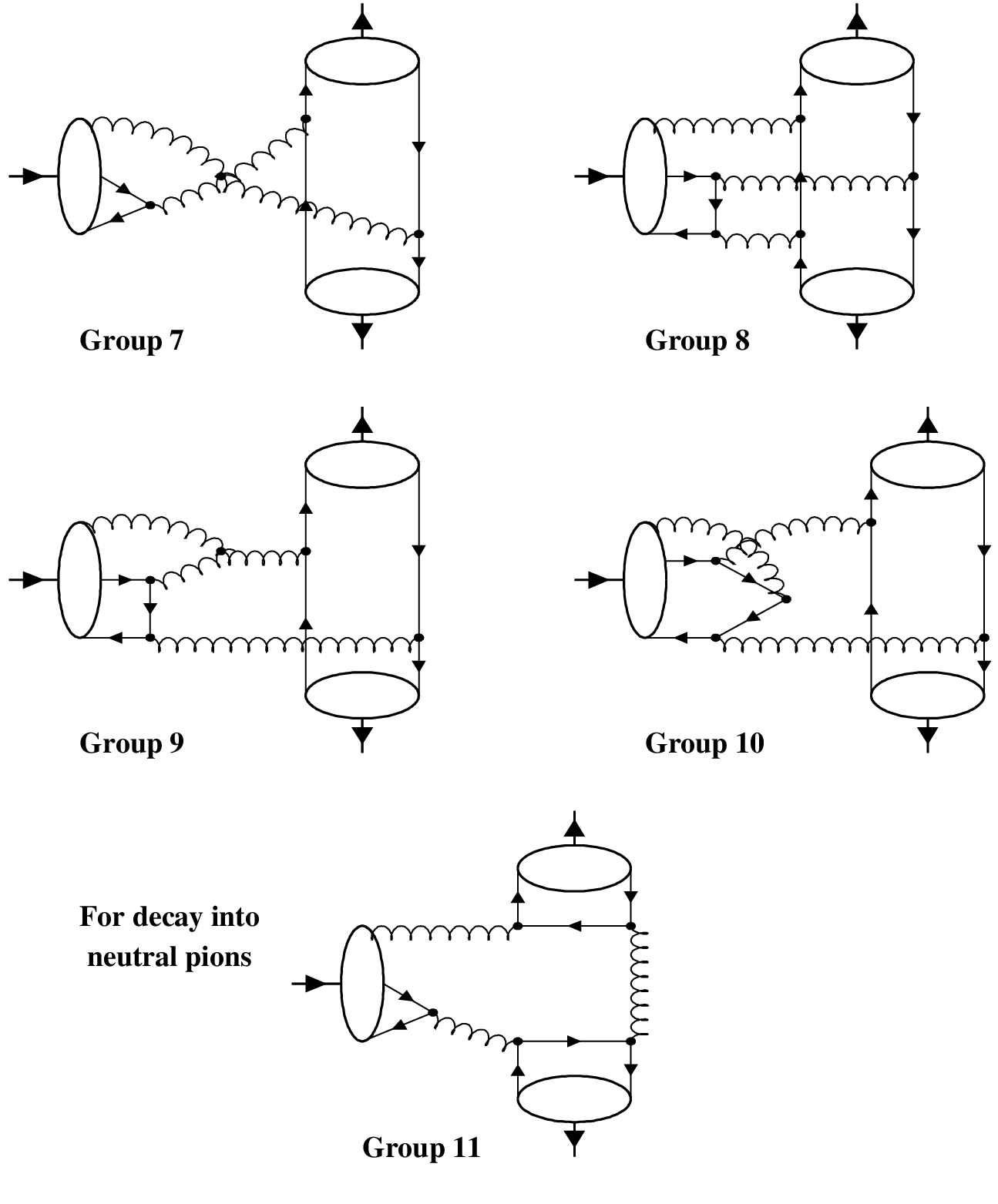,width=3.70cm}}
\caption{The groups of graphs that contribute to the decay into $\p\p$.}
\label{f:pi}
\efi
It remains to convolute the $\f_h$'s in the sHSA or $\j_h$ in the 
mHSA and $T_H$ together to get the partial widths for the different
two-body channels. 
\bfi[t]
\centerline{\epsfig{file=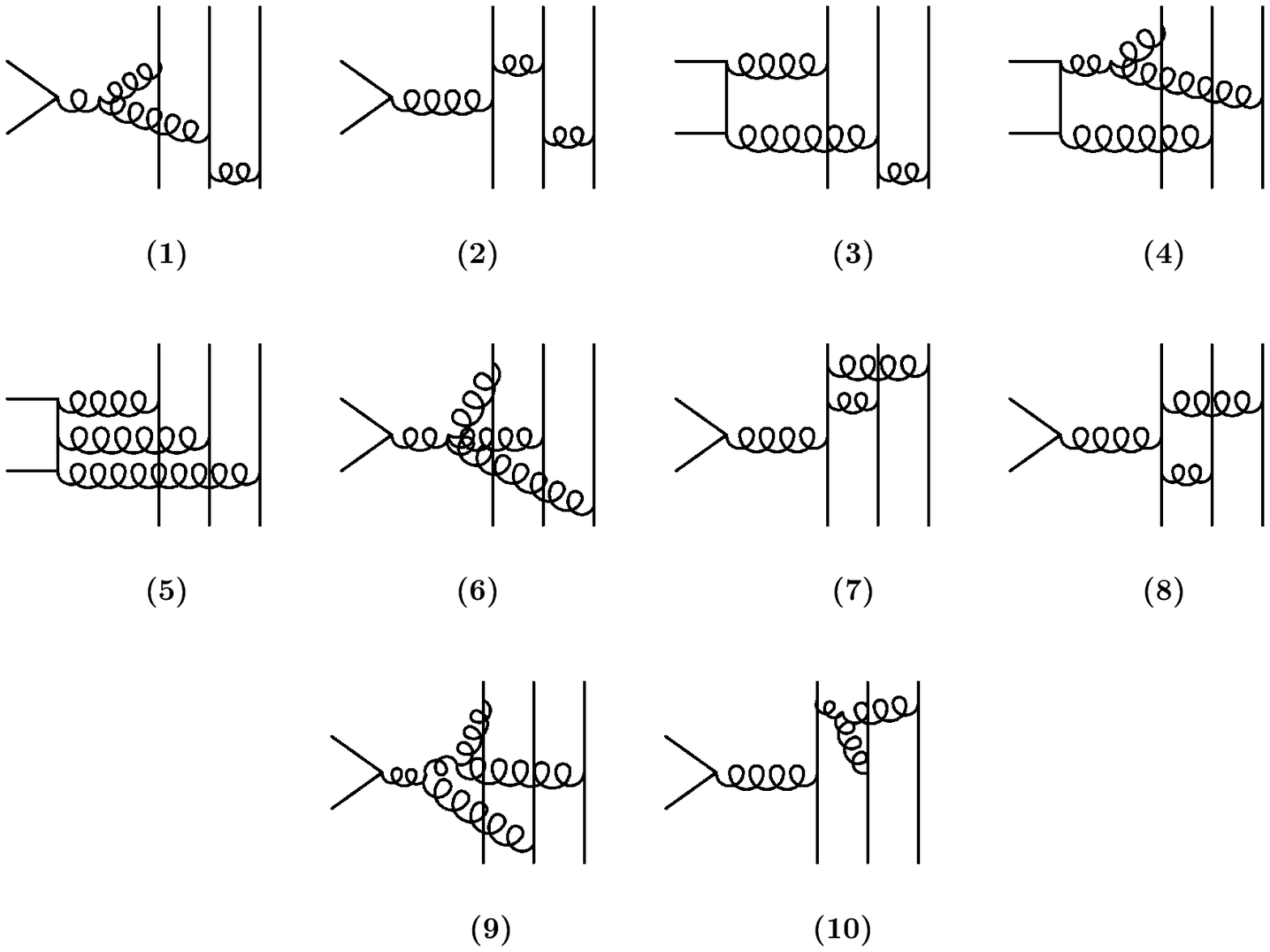,width=7.0cm}}
\caption{The groups that contribute to the decay into $p\bar p$. The
constituent gluon is not drawn here.}
\label{f:p}
\efi
The details for the calculation of the $\c_J$ decay into $\p\p$
are to be found in \cite{bks,bks2} and those for $\c_J$ decay into
$p\bar p$ are in \cite{w1,w3}.

\section{Combining color singlet and octet contributions}
\label{s:cs+co}

The arguments in Sec. \ref{s:lmd} show that for P-wave charmonium,
color octet and singlet are of equal weight and must both be included. 
With the calculation of the singlet contribution in Sec. \ref{s:2bd}
and the octet contribution in Sec. \ref{s:co}, the two can now be
combined to give, from a theoretical point of view, the proper
partial widths of $\c_J$. The results are tabulated in \tref{t:r3}
\footnote{These results for the decay into proton-antiproton are
final. They supersede all previously reported preliminary values 
in \cite{w2,w4}.}. 
Notice that the experimental measurements are different between the
PDG and BES data. It is up to the experimentalists to reconcile this
in the future. The important point here is that by including the
octet contributions, the theoretical widths are in much better shape
for explaining the measured data and also in accordance with theoretical 
expectation. With hindsight it is not entirely surprising that
color octet is required in the two-body decay channels since in 
inclusive decays it is needed already at the leading order for 
$\c_1$ and next-to-leading order for $\c_0$ and $\c_2$.  
\begin{flushleft}
\begin{table}
\caption{Singlet and singlet-octet combined partial widths.}
\begin{tabular}{ccccc} \hline
 $J$  & \multicolumn{4}{c}{$\Gamma_{{\c_J\rightarrow \pi\pi}}$ [keV]} 
                                                                \\ \cline{2-5}
      & $\Gamma^{(1)}$ & $\Gamma^{(1+8)}$ & PDG 
      & BES \                                                   \\ \hline
  0   &   8.22      &  45.4        & 105.0 $\pm$ 30   
      &  64.0 $\pm$ 21.0                                        \\
  2   &   0.41      &   3.64       &   3.8 $\pm$ 2.0  
      &   3.04 $\pm$ 0.73                                       \\ \hline 
\end{tabular} 
\begin{tabular}{ccccc} \hline
 $J$  & \multicolumn{4}{c}{$\Gamma_{{\c_J\rightarrow p\bar p}}$ [eV]} 
                                                                \\ \cline{2-5}
      & $\Gamma^{(1)}$ & $\Gamma^{(1+8)}$ & PDG  
      & \ BES \ \ \                                             \\ \hline
  1   &    3.15      &  56.27          &  75.68 $\pm$ 10.5  
      &   37.84                                                 \\
  2   &   12.29      &  154.19         & 200.00 $\pm$ 20.0  
      &  118.00                                                 \\ \hline 
\end{tabular} 
\label{t:r3}
\end{table}
\end{flushleft}

\section{Summary}
\label{s:sum}

In this talk, we showed with new calculations that valence
color singlet state of P-wave charmonia alone is insufficient
to account for experimental measurements of the partial decay 
widths in contradiction to those done in the past pre-dating the
emergence of the color octet state in quarkonium physics. 
Normally higher states should be suppressed but this was shown 
to be circumvented because of angular momentum suppression of 
the valence state with $L=1$ which brought it down to the same level 
of the octet state. Power dependence in the heavy charmonium mass 
arguments revealed that it was inconsistent to perform P-wave charmonium
calculations without including the color octet. This was eventually 
supported by explicit calculations. The anormaly of the situation in 
charmonium physics between inclusive and exclusive processes pointed 
out was thus reconciled. The arguments given here can be straight 
forwardly generalized to even higher-wave quarkonia. Calculations 
involving the latter require the inclusion of not only 
\newpage
\noindent
the next higher states but also those above them. Therefore one has 
to be very careful when dealing with quarkonia whose valence states 
have non-zero orbital angular momentum.

\end{document}